# Non-Reciprocal Willis Coupling in Zero-Index Moving Media

Li Quan[1], Dimitrios L. Sounas[1,2], and Andrea Alù[1,3,4,5,*]

[1]Department of Electrical and Computer Engineering, The University of Texas at Austin, Austin, TX 78712, USA

[2]Department of Electrical and Computer Engineering, Wayne State University, Detroit, MI 48202, USA

[3]Photonics Initiative, Advanced Science Research Center, City University of New York, New York, NY 10031, USA

[4]Physics Program, Graduate Center, City University of New York, New York, NY 10016, USA

[5]Department of Electrical Engineering, City College of The City University of New York, NY 10031, USA

[*]Corresponding author: aalu@gc.cuny.edu

**Mechanical motion can break the symmetry in which sound travels in a medium, but significant non-reciprocity is typically achieved only for very large motion speeds. Here we combine moving media with zero-index acoustic propagation, yielding extreme non-reciprocity and induced bianisotropy for modest applied speeds. The metamaterial is formed by an array of waveguides loaded by Helmholtz resonators, and it exhibits opposite signs of the refractive index sustained by asymmetric Willis coupling for propagation in opposite directions. We use this response to design a non-reciprocal acoustic lens focusing only when excitation from one side, with applications for imaging and ultrasound technology.**

Reciprocity in wave propagation demands that the response of a source remains the same when source and observation points are interchanged. Breaking this symmetry allows designing devices that exhibit different transmission for opposite propagation directions, which is important for protection of sensitive equipment from external interference and for full-duplex communications. For electromagnetic waves, several approaches have been proposed to break reciprocity, including biasing with external magnetic fields [1]-[2] transistors [3]-[4], angular momentum [5]-[6], spatiotemporal modulation [7]-[8], and non-linearities [9]-[10]. In acoustics, non-reciprocal devices have been mainly realized based on nonlinear mechanisms [11]-[13]. It is well known that sound traveling parallel or anti-parallel to a moving medium is transmitted non-reciprocally [14], however strong effects are typically achieved only when the velocity of the medium is large, or in highly resonant devices. Based on this principle, momentum bias applied through moving media was recently used to realize linear acoustic non-reciprocal devices [15]-[16]. In the following, we explore moving metamaterials operated around their zero-index operation, showing how in this scenario mechanical motion opens highly unusual scenarios for sound propagation.

A moving medium exhibits different wave-vectors $k_+$ and $k_-$ for opposite propagation directions, which is a signature of nonreciprocity. As such, the nonreciprocal coefficient

$$\eta = \left|\text{Re}(k_+) + \text{Re}(k_-)\right| / \left|\text{Re}(k_+) - \text{Re}(k_-)\right| \tag{1}$$

measures the degree of asymmetry in wave propagation for opposite directions. In absence of motion, $k_+$ and $k_-$ are necessarily the same in magnitude and with opposite signs, leading to $\eta = 0$, while $\eta$ is nonzero when reciprocity is broken. In a moving non-dispersive medium, for small velocities $k_{\pm} = \pm k_R + k_{NR}$ [17], where $k_R = k_0/(1-M^2)$ and $k_{NR} = -Mk_0/(1-M^2)$ are the reciprocal and non-reciprocal portions of the wavenumber, respectively, $M = U_0/c_0$ is the Mach number, defined as the ratio of flow speed $U_0$ to the background sound speed $c_0$, and $k_0 = \omega/c_0$ is the wavenumber in free space. Replacing these quantities in (1) yields $\eta = |k_{NR}/k_R| = |M|$, implying that non-negligible non-reciprocity can be expected only for large, often impractical Mach numbers.

In order to break this trade-off and achieve large non-reciprocity with small flow speed, we explore the regime for which $k_R = 0$, i.e., zero-index propagation, so that the non-reciprocal portion of the wave-number $k_{NR}$ dominates. In electromagnetics, epsilon-near-zero (ENZ) metamaterials [18]-[19] provide a reciprocal zero index of refraction, which has been shown to lead to extreme wave propagation properties [20]-[21]. Acoustic waves in a zero-index media may therefore provide a platform to boost nonreciprocal phenomena when modest medium speeds are considered. Density-near-zero metamaterials [22]-[23], the analogue of ENZ materials for sound, have been realized in the past using waveguides loaded by membranes. However, this approach is not suitable for our purpose, because membranes block material flow. We consider therefore extreme non-reciprocal responses by imparting air flow to waveguides loaded by a Helmholtz resonator array, inducing a near-zero refractive index at rest ($k_R \approx 0$) [24], while $k_{NR} \neq 0$. In such a metamaterial, we expect that even a small mechanical motion yields a negative phase velocity (and refractive index) for propagation parallel to the fluid motion, and a positive one for propagation anti-parallel to it, hence $\eta = |k_{NR}/k_R| \to \infty \gg |M|$. In the following, we verify this response for plane-wave incidence, observing opposite refraction angles for excitation from opposite

sides. After explaining the physics around this unusual response, and its relation to non-reciprocal Willis coupling, we then utilize these properties to design a planar non-reciprocal lens, which can focus a source from one side, but acting as a divergent lens for opposite excitation.

The geometry under analysis is shown in Fig. 1a and it consists of an array of parallel waveguides loaded with Helmholtz resonators. The green color indicates the region with air flow, and the geometrical parameters are provided in the caption. In order to impart air flow in each waveguide, a pipe is connected to both sides and loaded with a fan, (not shown in the figure), inducing a continuous air flow, as indicated by the arrows. To ensure that the incident acoustic wave only travels through the waveguide, we load two Helmholtz resonators at each pipe opening (not shown in the figure), designed to filter out the frequencies of excitation. The structure is supplied with two quarter-wavelength matching layers at both sides (yellow color), in order to eliminate impedance mismatch. Figure 1b shows the calculated Mach number along each waveguide, which starts at zero outside the moving segment, and it linearly increases to an approximately constant value of 0.1 inside the waveguide through a finite transition layer. The small fluctuations of the Mach number inside the waveguide are due to the presence of the Helmholtz resonators. The numerical simulations here and in the following are performed using COMSOL Multiphysics [25].

Consider now the excitation of this geometry with obliquely-incident plane waves from opposite sides, as shown in Fig. 2. In our simulations, we use the continuity boundary conditions for acoustic pressure and particle velocity at the interface between the region with no motion and the matching layer. At the interface between the region with no motion and the transition layer inside the waveguides, as well as at the one between the transition layer and the main body of the waveguide, we instead use continuity boundary conditions for the acoustic pressure and the air mass flow [26]. Figure 2a presents the acoustic pressure field distribution for excitation from the left side. Due to momentum conservation, the tangential component of the wave-vector is the same everywhere in space, while the normal component changes direction across the interface, resulting in negative refraction. Quite interestingly, the situation is opposite for excitation from the right side: Figure 2(b) presents the acoustic fields for an incident wave coming from the right side at the same incident angle. In this case, the direction of the normal component of the wave-vector does not flip as the wave enters the metamaterial, indicating positive refraction. This drastically different refraction response from opposite sides is a signature of extreme nonreciprocity, arising from the moving medium combined with the index-near-zero response in the metamaterial. We stress that this is achieved for modest values of $M$.

A better understanding of the phenomenon can be gained by extracting the effective constitutive parameters of the metamaterial. The mass conservation equation and momentum equation in each waveguide with air flow are derived in detail in [27], and take the general form

$$\partial u/\partial x = i\omega\left(E_{eff}^{-1} p + \xi_{eff} u\right) \quad (2)$$

$$\partial p/\partial x = i\omega\left(\rho_{eff} u + \varsigma_{eff} p\right), \quad (3)$$

where $E_{eff}^{-1} = \left(E_0^{-1} - F^{-1}\right)/\left(1-M^2\right)$ is the effective bulk modulus, $\rho_{eff} = \rho_0/\left(1-M^2\right)$ is the effective density, $E_0 = \rho_0 c_0^2$ is the bulk modulus in air, $\xi_{eff} = -M c_0^{-1}/\left(1-M^2\right)$ and $\varsigma_{eff} = M\left(\rho_0 c_0 F^{-1} - c_0^{-1}\right)/\left(1-M^2\right)$ are odd bianisotropy cross coupling coefficients arising from non-reciprocity, $F = S_w dM_a\left(\omega^2 - \omega_0^2\right)$ is a factor that depends on the geometry of the structure, and $\omega_0$, $M_a$ are the Helmholtz resonator resonance angular frequency and acoustical mass, respectively. The dispersion of the effective parameters versus frequency for the metamaterial geometry at hand is shown in [27].

The mechanical motion in each waveguide introduces highly unusual propagation properties around the zero-index propagation regime. For a stationary waveguide loaded with Helmholtz resonators [28], which corresponds to our scenario in the limit of $M = 0$, only the bulk modulus $E_{eff}$ is affected by the loads, yielding near-zero $E_{eff}$, corresponding to a zero index of refraction, when $E_0^{-1} = F^{-1}$. The effective density is not affected by the resonators. When a modest fluid motion is considered, the effective density and bulk modulus are weakly modified through the factor $(1-M^2)^{-1}$, but most importantly they are coupled together through the bianisotropy coefficients $\xi_{eff}$ and $\varsigma_{eff}$, also known as Willis coupling, which is at the core of the described phenomena. Different from conventional Willis coupling [29]-[32], these coefficients do not obey reciprocity, $\xi_{eff} \neq -\varsigma_{eff}$, and are odd with respect to $M$, i.e., they flip sign for opposite propagation directions, a clear sign of non-reciprocity. As we show in [27], around the zero-index operation $\varsigma_{eff}$ goes through a resonance and flips sign, similar to $E_{eff}$, producing extremely asymmetric Willis coupling coefficients and non-reciprocal response at the zero-index operation. By combining Eqs. (2) and (3), we derive the dispersion relation

$$k_{\pm} = \pm \frac{\omega\sqrt{(\xi-\varsigma)^2 + 4\rho_{eff} E_{eff}^{-1}}}{2} + \frac{\omega(\xi+\varsigma)}{2}, \quad (4)$$

yielding $k_R = \omega\sqrt{(\xi-\varsigma)^2 + 4\rho_{eff} E_{eff}^{-1}}/2$ and $k_{NR} = \omega(\xi+\varsigma)/2$. At the angular frequency

$\omega = \sqrt{\omega_0^2 + \left(1 + \sqrt{1-M^2}\right) E_0 \big/ \left(2 S_w d M_a\right)}$, $k_R = 0$, and $\eta \to \infty$, largely enhancing the non-reciprocal response, due to asymmetric Willis coupling induced by the mechanical motion. In this regime the wavenumber reads

$$k_\pm = k_{NR} = -k_0 \frac{M}{\left(1 + \sqrt{1-M^2}\right)\sqrt{1-M^2}}, \quad (5)$$

enabling opposite refractive index for opposite directions of propagation. Interestingly, the wavenumber in the metamaterial has the same real value for both propagation directions, i.e., no matter whether the incident wave is coming from left or right, the wavevector has the same direction and value, anti-parallel to the motion, consistent with our numerical simulations in Fig. 2. We stress that for $k_R = 0$ the effective bulk modulus is negative, $E_{eff}^{-1} = -\left(\xi_{eff} - \varsigma_{eff}\right)^2 \big/ \left(4 \rho_{eff}\right)$, and the effective density positive, $\rho_{eff} = \rho_0 / \left(1 - M^2\right)$, yet the acoustic wave travels in the metamaterial without decay because of the strong Willis coupling response.

Figure 3 shows the dispersion of the wavenumber in Eq. (4). In absence of air flow ($M$ = 0), the dispersion has a cut-off at the zero-index condition $E_0^{-1} = F^{-1}$, and it is strictly even with respect to $k$, as expected from reciprocity. When a moderate air-flow is turned on ($M$ = 0.1), the dispersion diagram is asymmetric, as expected for a nonreciprocal medium, and the cutoff frequency shifts down to $\omega = \sqrt{\omega_0^2 + \left(1 + \sqrt{1-M^2}\right) E_0 \big/ \left(2 S_w d M_a\right)}$. Around this frequency, waves propagating in opposite directions have a nonzero (negative) wavenumber, independent of the propagation direction, and the non-reciprocity coefficient $\eta$ is very large.

The non-reciprocal Willis coupling introduced here through mechanical motion at the zero-index frequency can be used to create a lens that focuses a source placed at one side, but with diverging properties when a source is placed on the other side. The focusing operation is achieved by imparting a phase shift across the structure that transforms a diverging circular wavefront to a converging one [27], which is achieved tailoring the air flow velocity across different channels to accumulate the required phase at each aperture. Our design is shown in Fig. 4(a), where we plot the relation between Mach number, maintained small in each channel, and the channel number $n$, with $n = 0$ being the channel on the same axis as the source ($y = 0$). The air flow in each channel is symmetric with respect to the $y$-axis (i.e., for channels $N$ and $-N$ the air flow is the same), so we only show the imparted Mach number for positive $n$. Figure 4(b) presents the calculated acoustic pressure distribution when the sound source is located on the left of the lens: after travelling through the planar metamaterial, a focused image is constructed at the right side of the lens. Figure 4(c) presents instead the pressure distribution when the sound source is located at the right side of the lens: here we only get a divergent wave, demonstrating strong nonreciprocity with modest required flow velocities, enabled by operating around the zero-index frequency of the metamaterial.

By adjusting the air flow in each channel, we can get different functionalities. In Figs. 4d-f we show a design that converts a point source to a plane wave only when the source is located on the right side of the metamaterial. Again, the relation between channel number and Mach number is presented in Fig. 4(d), while Figs. 4e-f show the acoustic pressure profile when the source is located at the two sides of the structure [27], highlighting again the strongly non-reciprocal response.

In conclusion, we have presented here a Willis metamaterial operating near the zero-index operation, yielding extreme non-reciprocal responses with modest air flow. These unusual acoustic properties are the result of a non-reciprocal bianisotropic response dominating the effective properties of the metamaterial, as the reciprocal response tends to zero. For this reason, even modest air flows can provide highly unusual responses, including opposite (positive-to-negative) refractive index for opposite propagation directions, and non-reciprocal lenses. The air flow can be modulated in real-time to reconfigure the properties of the metamaterial, making it an exciting platform to control sound beyond the conventional limitations of natural materials. We envision a plethora of applications of these concepts, from ultrasound imaging to sonar technology, with possible extensions to the realm of phononics and surface acoustic waves.

**Figures**

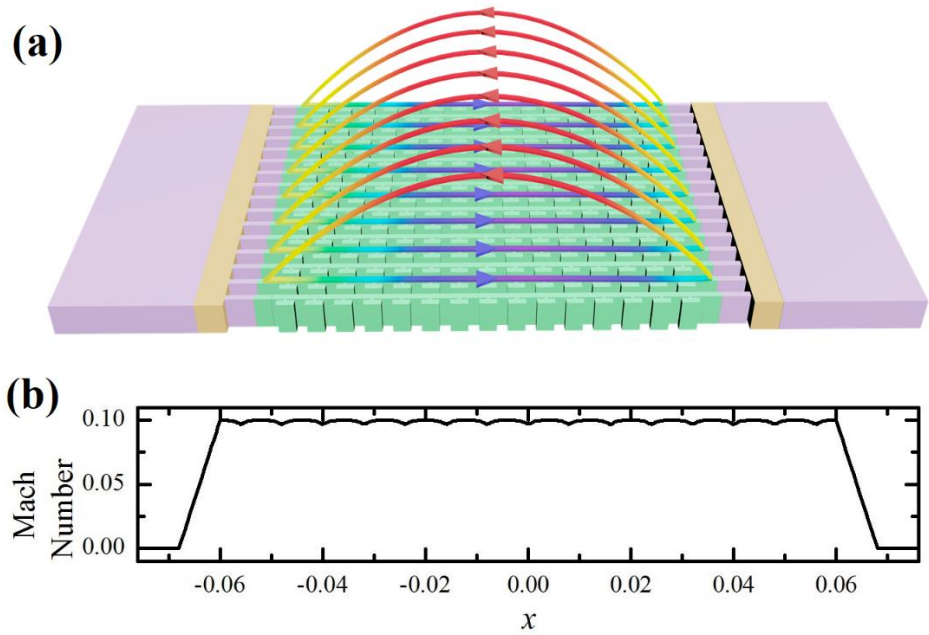

Fig. 1. (a) Geometry of the nonreciprocal metamaterial, formed by an array of parallel waveguides loaded by Helmholtz resonators (green). A constant air flow inside the waveguides is generated by fans. The waveguides have a width $S_w$=3 mm, a distance $d$=8 mm between neighboring resonators, which have a neck length $l$=0.5 mm, a neck width $a$=1 mm, a cavity length $b$=7 mm and a cavity width $h$=3.5 mm. (b) Mach number distribution in the waveguide between two matching layers.

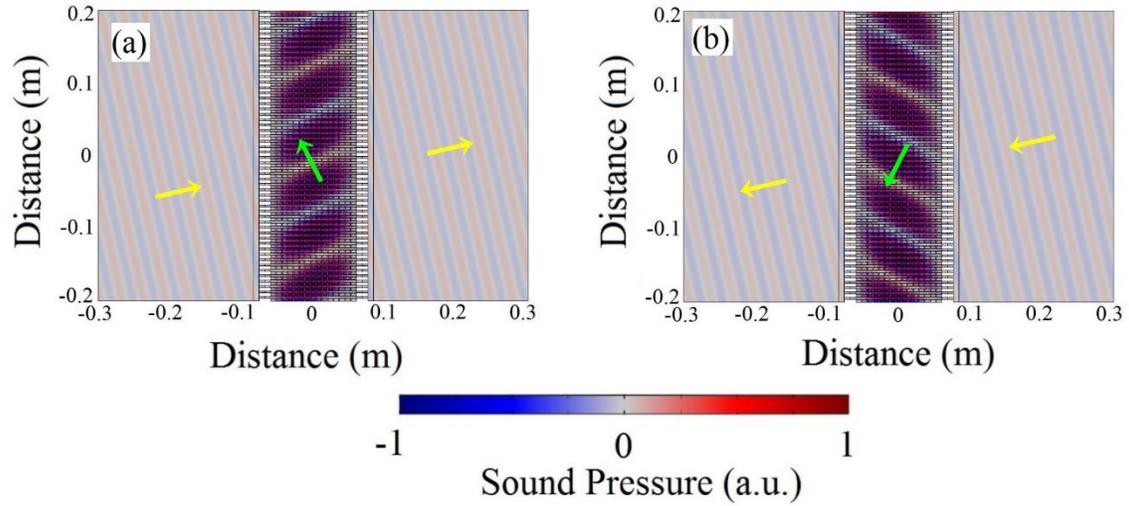

Fig. 2. (a) Acoustic pressure distribution for an incident wave coming from the left. The yellow arrows indicate the wave vector in air and the green arrow indicates the wave vector in the metamaterial. (b) Acoustic pressure distribution for an incident wave coming from the right.

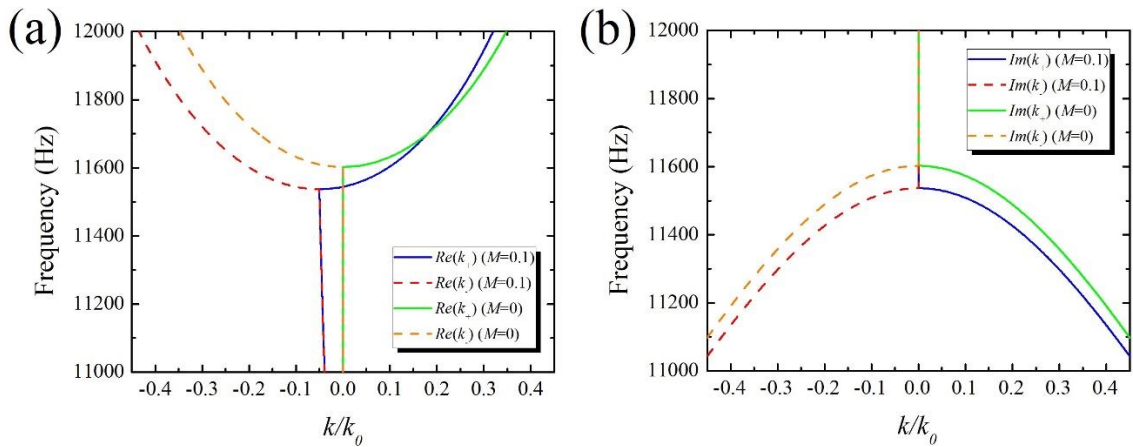

Fig. 3. Dispersion diagram for the geometry of Fig. 1(a). (a) Real part. (b) Imaginary part.

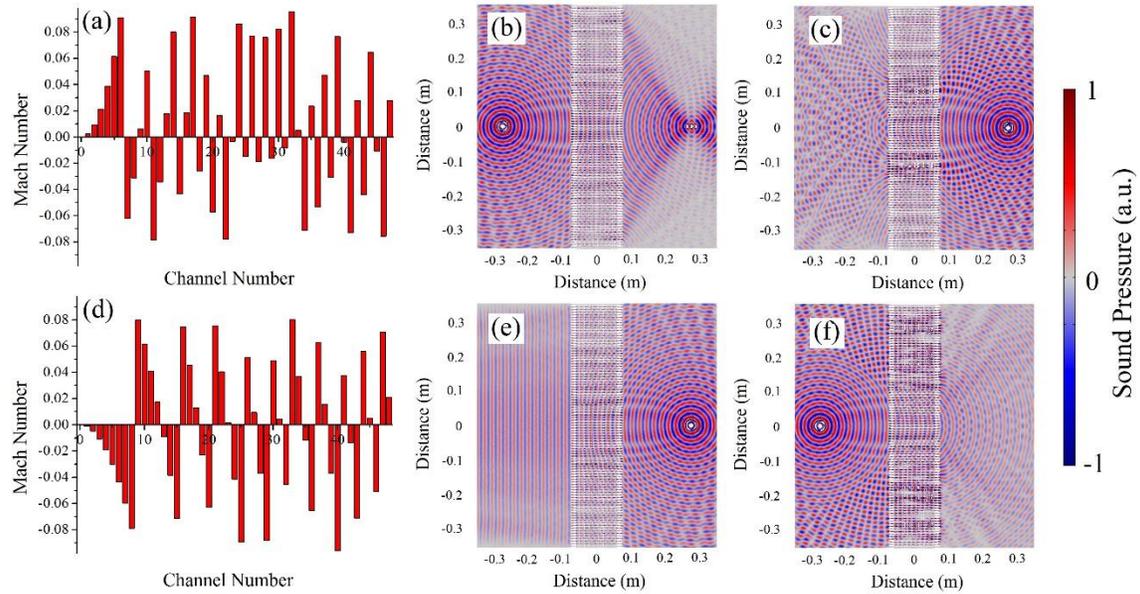

Fig. 4. (a) Modulation of the Mach number in each channel to synthesize a focusing lens. (b) Acoustic pressure distribution when the source is located on the left side of the lens. A focused image is obtained on the right. (c) Acoustic pressure distribution when the source is located on the right of the lens. (d) Modulation of the Mach number to synthesize a point-source to plane wave converter. (e) Acoustic pressure distribution when the source is located on the right. A plane wave is induced on the left side. (f) Acoustic pressure distribution when the source is located on the left side.